\documentstyle[aps,epsfig]{revtex}
\begin{document}
\draft
\preprint{ }
\title{Vortices in Bose-Einstein-Condensed Atomic Clouds}
\author{Emil Lundh$^{1,2}$,  C.\ J.\ Pethick$^{2,3}$ and H.\ Smith$^{4}$}
\address{$^1$Department of Theoretical Physics,  Ume{\aa} University,
S-90187 Ume{\aa}, Sweden
\\
$^2$Nordita, Blegdamsvej 17, DK-2100 Copenhagen \O, Denmark\\
$^3$Department of Physics, University of Illinois at Urbana-Champaign,
1110 W.\ Green Street, Urbana, Illinois 61801\\
$^4$\O rsted Laboratory, H.\ C.\ \O rsted Institute,
Universitetsparken 5, DK-2100 Copenhagen \O, Denmark\\}

\date{\today}

\maketitle

\begin{abstract}
The properties of vortex states in
a  Bose-Einstein condensed cloud of atoms are considered at zero temperature.
Using both
 analytical and numerical methods we solve  the
time-dependent Gross-Pitaevskii equation for the
case when a cloud of atoms containing a vortex is released from a trap.
In two dimensions we find the simple result that
the time dependence of the cloud radius is  given by
$(1+\omega^2t^2)^{1/2}$, where $\omega$ is the trap frequency.
We  calculate and compare
 the expansion of the vortex core and the cloud radius
for   different numbers of particles and interaction
strengths, in both two and three dimensions,
and discuss the circumstances under which vortex states may be observed
experimentally.

   \end{abstract}


\section{Introduction}

Interest in the properties of  atomic clouds
was greatly stimulated by the experimental discovery\cite{jila1,rice,mit1} of
Bose-Einstein condensation in  trapped gases of alkali-metal atoms.
One of the intriguing features of these novel condensates that
remains to be explored experimentally
is their behavior under rotation.

Vortex states in trapped atomic clouds at zero temperature
have been considered  theoretically
 by several authors \cite{dalfovo,lps,rokhsar}. From numerical
solutions to the Gross-Pitaevskii equation
Dalfovo and Stringari \cite{dalfovo} determined the
 critical angular velocity, that is the lowest angular velocity for
which it is favorable for a vortex to enter the cloud.
 Lundh {\it et al.} \cite{lps}
obtained for reasonably large clouds approximate analytical expressions 
for the critical
angular velocity which agree closely with the 
numerical results.

For large clouds the presence of a vortex is difficult to detect
experimentally,
since the size of the vortex core is small compared to the size of the cloud.
Consequently, the energy of a state in which a vortex is present
is nearly  equal to the energy of the ground
state without the vortex. Rather than measuring the total energy
of a trapped cloud it may therefore
be advantageous to investigate the density profile of the cloud during a free
expansion
after the trap potential has been turned off. In this way one
may be able to follow  how the ``hole'' in the middle of the cloud
develops as a function of time, thereby  allowing one to distinguish
the free expansion of the vortex state from that of  the ground state
without a vortex.
In this paper we  shall therefore study  the free
expansion of a cloud containing a vortex, which is initially
trapped in an anisotropic harmonic-oscillator potential
and subsequently released.

Apart from the difficulties involved in the detection of a vortex
state, one also has to consider its experimental generation \cite{mzh}.
Due to the energy barriers separating rotating and non-rotating states
it may  be advantageous to generate a vortex state by cooling a rotating
cloud in its normal state below the transition temperature rather
than rotating the cloud from rest at temperatures
below the Bose-Einstein condensation temperature.

There are questions regarding the stability of vortices \cite{rokhsar}, and the
 character of the ground state for a fixed angular momentum if the angular
momentum per particle is not a multiple of $\hbar$ \cite{ho}.  
However, we shall
not address these issues in this paper, but will consider the case of expansion
of a cloud containing a vortex with one quantum of circulation.

In the following we shall  consider the free expansion of
a  condensed atomic cloud in the limit when the
temperature  is sufficiently low that the influence of
the normal component  is negligible. We
assume that the system is  dilute, in the sense  that
the scattering length is much less than the
interparticle distance. In this case one
may neglect the depletion of the condensate,
and the wave function, $\psi(\vec{r},t)$,   of
the condensed state in an external potential $V(\vec{r}\,)$
satisfies the time-dependent Gross-Pitaevskii (GP) equation \cite{gp}
\begin{eqnarray}
 i\hbar\frac{\partial \psi(\vec{r},t)}{\partial t}=  \left[
   -\frac{\hbar^2}{2m} \nabla^2
   + V({\vec r})
    +U_0|\psi(\vec{r})|^2\right]\psi(\vec{r}).
\label{gp}
\end{eqnarray}
 The effective two-body interaction  $U_0$
is given by $U_0=4 \pi \hbar^2 a/m$,
where $a$ is the scattering length and $m$ is
the atomic mass. We shall assume the scattering length to be positive,
thereby ensuring the stability of the non-rotating 
condensed state for any value
of the total number of particles, $N$.

    The plan of the paper is as follows:  In Sec.\ II we consider a
two-dimensional vortex state and show that
its development in time is given approximately
 by a simple,  analytical expression for the cloud radius.
We also solve the time-dependent Gross-Pitaevskii equation
 numerically, and compare the resulting density profile to
 the one obtained analytically. In Sec.\ III we discuss the
more realistic three-dimensional case by an approximate variational
method as well as by  numerical methods. The expansion
of the rotating cloud of  atoms is compared to that obtained
for a non-rotating cloud when the trap potential is turned off.
Section IV is a brief conclusion.

\section{Vortices in two dimensions}

In the following we consider a two-dimensional geometry, where
the vortex is uniform along its axis, which we take to be the $z$-axis.
The harmonic-oscillator potential is assumed to be isotropic in the
$xy$-plane and given by
\begin{equation}
V=\frac{1}{2}m\omega^2\rho^2,
\end{equation}
where $\rho^2=x^2+y^2$.

In the stationary case the condensate wave function depends on time
only through its phase according to
\begin{equation}
\psi({\vec r},t) =\psi({\vec r}\,)e^{-i\mu t/\hbar},
\label{my}
\end{equation}
where $\mu$ is the chemical potential. In the Thomas-Fermi
approximation the  chemical potential is obtained by
inserting Eq.\ (\ref{my}) into Eq.\ (\ref{gp}) and neglecting the
kinetic-energy term. This results in
the following expression for the particle density:
\begin{equation}
|\psi|^2=\frac{1}{U_0}(\mu-\frac{1}{2}m\omega^2\rho^2),
\end{equation}
provided $\rho$ is less than $\rho_{\mathrm max}=(2\mu/m\omega^2)^{1/2}$.
For $\rho$ greater than $\rho_{\mathrm max}$, the density
is zero.

The number of particles, $\nu$, per unit length along the $z$-axis
is
\begin{equation}
\nu=2\pi\int_0^{\rho_{\mathrm max}}d\rho\,
\rho|\psi|^2=\frac{\pi\mu^2}{U_0m\omega^2}.
\end{equation}
In terms of the dimensionless parameter $\gamma$, which is defined by
\begin{equation}
\gamma=\nu a,
\end{equation}
the chemical potential is then seen to be given by
\begin{equation}
\mu =2\hbar\omega\gamma^{1/2}.
\label{mures}
\end{equation}
The Thomas-Fermi approximation becomes exact in the limit of large
$\gamma$. Since the chemical potential is
given in terms of the total energy per unit length, $E$,
 by the equation $\mu=\partial E/\partial \nu$, we conclude that
$\mu=3E/2\nu,$
and therefore
\begin{equation}
E=\frac{4}{3}\nu\hbar\omega\gamma^{1/2}.
\label{eres}
\end{equation}

Since we shall use Gaussian trial functions later on in
discussing the time evolution of the vortex state, it is instructive
to compare Eqs.\ (\ref{mures}) and (\ref{eres}) with the result
of calculating the ground state energy variationally with a
trial function of the form
\begin{equation}
\psi(x,y)= Ae^{-\rho^2/2b^2}
\end{equation}
where $b$ is a variational parameter. Writing $b=\alpha a_{\mathrm osc}$,
where $a_{\mathrm osc} = (\hbar/m\omega)^{1/2}$ is the oscillator length,
one finds
\begin{equation}
E=\nu\hbar\omega(\frac{1}{2\alpha^2}+
\frac{1}{2}\alpha^2 + \frac{\gamma}{\alpha^2}).
\end{equation}
This expression has a minimum  for
$\alpha=\alpha_0=  (1+2\gamma)^{1/4},$
and therefore the variational estimate of the energy of the ground state is
\begin{equation}
E=\nu\hbar\omega\sqrt{2\gamma+1}.
\label{varground}
\end{equation}
In the limit of large $\gamma$ this variational approximation to
the  energy exceeds the asymptotically exact result (\ref{eres}) by only 6\%.
In the opposite limit, for small $\gamma$, it is exact  to
order $\gamma$.

We now  consider a vortex state
corresponding to one quantum $h/m$
of circulation around the $z$-axis. The corrections to the Thomas-Fermi
energy (\ref{eres}) consists of two terms: the vortex energy
per unit length, and the kinetic energy associated with
the rounding of the density profile in the vicinity of
the surface of the cloud, which is of the same order of magnitude
as the vortex energy. For large clouds the vortex energy per unit length  
is\cite{lps}
\begin{equation}
E_v = \pi |\psi(0)|^2\frac{\hbar^2}{m}\ln \left(
   \frac{0.888 \rho_{\mathrm max}}{\xi_0} \right)\approx \frac{\nu \hbar
\omega}{2 \gamma^{1/2}}\ln (3.55 \gamma^{1/2}),
\label{vortex}
\end{equation}
where the coherence length $\xi_0$ is defined by
\begin{equation}
 \frac{\hbar^2}{2m\xi_0^2}=U_0|\psi(0)|^2.
\label{coherencelength}
\end{equation}
Note that the ratio of the coherence length to the cloud radius in
the Thomas-Fermi limit is given by
\begin{equation}
\frac{\xi_0}{\rho_{\mathrm max}}=\frac{\hbar\omega}{2\mu}
= \frac{1}{4 \gamma^{1/2}}.
\end{equation}
The kinetic energy associated
with the surface is\cite{lps}
\begin{equation}
E_k = \nu\frac{\hbar\omega}{4\gamma^{1/2}} \ln(2.43\gamma^{1/3}).
\end{equation}
Summing up the various contributions, the total energy per unit length
for the cloud in the vortex state is therefore
\begin{equation}
E=\nu \hbar\omega[\frac{4}{3}\gamma^{1/2}
+\frac{1}{3\gamma^{1/2}}\ln (13.0 \gamma)]
\label{total}
\end{equation}
for sufficiently large clouds ($\gamma \gg 1$).
To leading order in $\gamma$ the energy of the vortex state is thus the same
as that of the ground state.
We may compare the result (\ref{total}) for the energy to the result of
using a variational
trial function of the form
\begin{equation}
\psi(\rho,\phi)=A\rho e^{i\phi} e^{-\rho^2/2b^2},
\label{trytry}
\end{equation}
where $\phi$ is the azimuthal angle, yielding
\begin{equation}
E=\nu\hbar\omega\sqrt{2\gamma+4}.
\label{varvort}
\end{equation}
The variational estimate of the vortex energy is obtained by subtracting
(\ref{varground}) from (\ref{varvort}), and gives the value
$\nu \hbar \omega 3\sqrt{2} / 4 \gamma^{1/2}$, which is to be compared with
Eq.\,(\ref{vortex}).
This shows that the
variational calculation gives a  poor estimate of the
energy of the vortex for strong coupling.  The
reason for this is that the trial
wave function has only one length scale, which determines both the size of the
cloud, and the size of the vortex core.

\subsection{Free expansion}

We shall now use Gaussian trial functions also for
treating the time development of the density,
when the cloud is released from the trap at a certain
instant of time,  $t=0$. We assume the solution to be homologous in the
sense that the local velocity is proportional to the distance from
the axis and therefore employ a trial function of
the form
\begin{equation}
\psi(\rho,\phi,t)=A\rho e^{i\phi} e^{-\rho^2/2R^2}e^{i\beta\rho^2/2},
\end{equation}
where $A=\sqrt{\nu/\pi}R^{-2}$ is the time-dependent normalization
constant found from conserving the number of particles per unit length.
The radial velocity is given by the derivative of the phase with
respect to $\rho$. The quantities
$R$ and $\beta$ depend on time.
We wish to evaluate the Lagrangian
\begin{equation}
\label{lagrangian}
L=\int_0^{\infty}d\rho 2\pi\rho\left(\frac{i\hbar}{2}
(\psi^*\frac{\partial\psi}{\partial t}-\psi\frac{\partial\psi^*}{\partial t})
-\frac{\hbar^2}{2m}|\nabla\psi|^2-\frac{U_0}{2}|\psi|^4\right)
\end{equation}
following the method used in Ref.\cite{perez}. Performing the
integration over $\rho$,
we obtain the Lagrangian as a function of the
variables $\beta$ and $R$ and their time derivatives $\dot{\beta}$ and
$\dot{R}$:
\begin{equation}
L=-\nu\left(\frac{\hbar^2}{mR^2}(1+\beta^2 R^4) +\nu\frac{U_0}{8\pi R^2}
+\dot{\beta}\hbar R^2\right).
\end{equation}
From the Lagrange equations for the two independent variables $\beta $ and
$R$ we obtain
\begin{equation}
\beta=\frac{m\dot{R}}{\hbar R},
\label{beta}
\end{equation}
and
\begin{equation}
\dot{\beta}=\frac{\hbar}{mR^4}-\frac{\hbar\beta^2}{m}
+\frac{\nu U_0}{8\pi\hbar R^4}.
\label{betadot}
\end{equation}
When Eq.\ (\ref{beta}) for $\beta$ is inserted in (\ref{betadot})
we arrive at the acceleration equation
\begin{equation}
\label{acceleration}
m\ddot{R}=-\frac{\partial U(R)}{\partial R},
\end{equation}
where
\begin{equation}
U(R)=\frac{\hbar^2}{2mR^2}(1+\frac{\gamma}{2}).
\end{equation}
Note that the contributions to the effective potential energy, $U$, from the
kinetic term due to the zero point motion, and the interaction
energy both scale
in the same way as a function of the size of the cloud.
This result is peculiar to two dimensions.
Equation (\ref{acceleration}) has the solution
\begin{equation}
R^2(t)= R^2(0) +v_0^2t^2,
\label{rad}
\end{equation}
where the velocity $v_0$ is given by
\begin{equation}
v_0^2=\frac{\hbar^2}{m^2R(0)^2}(1+\frac{\gamma}{2}).
\label{v0}
\end{equation}
Using the result of the minimization in the stationary case
with the trial function (\ref{trytry}), one finds
\begin{equation}
R(0)^4=a_{\mathrm osc}^4 (1+\frac{\gamma}{2}),
\end{equation}
and thus
\begin{equation}
v_0=a_{\rm osc}\omega(1+\frac{\gamma}{2})^{1/4}.
\label{v0final}
\end{equation}
This implies that the result (\ref{rad}) may be written
in the  form
\begin{equation}
R(t)^2= R(0)^2(1+\omega^2t^2).
\label{fin}
\end{equation}
This simple result is a consequence of the fact that the effective potential
energy varies as $R^{-2}$, which, as we remarked above,
is a special feature of two dimensions.

We note that the root-mean-square (rms) radius
$\rho_{\mathrm rms}$ is $\sqrt{2}R$,
so the final value of the rms velocity
is $v_{\mathrm rms}=\sqrt{2} v_0$.
Figure 1 shows how the final value of $v_{\mathrm rms}$
varies with
$\gamma$. For a non-rotating cloud, the corresponding analysis
is easily carried out with the result
$v_{\mathrm rms} = v_0 = a_{\mathrm osc} \omega (1+2\gamma)^{1/4}$.
This is included in Fig.\ 1.
\begin{figure}
\begin{center}
\epsfig{file=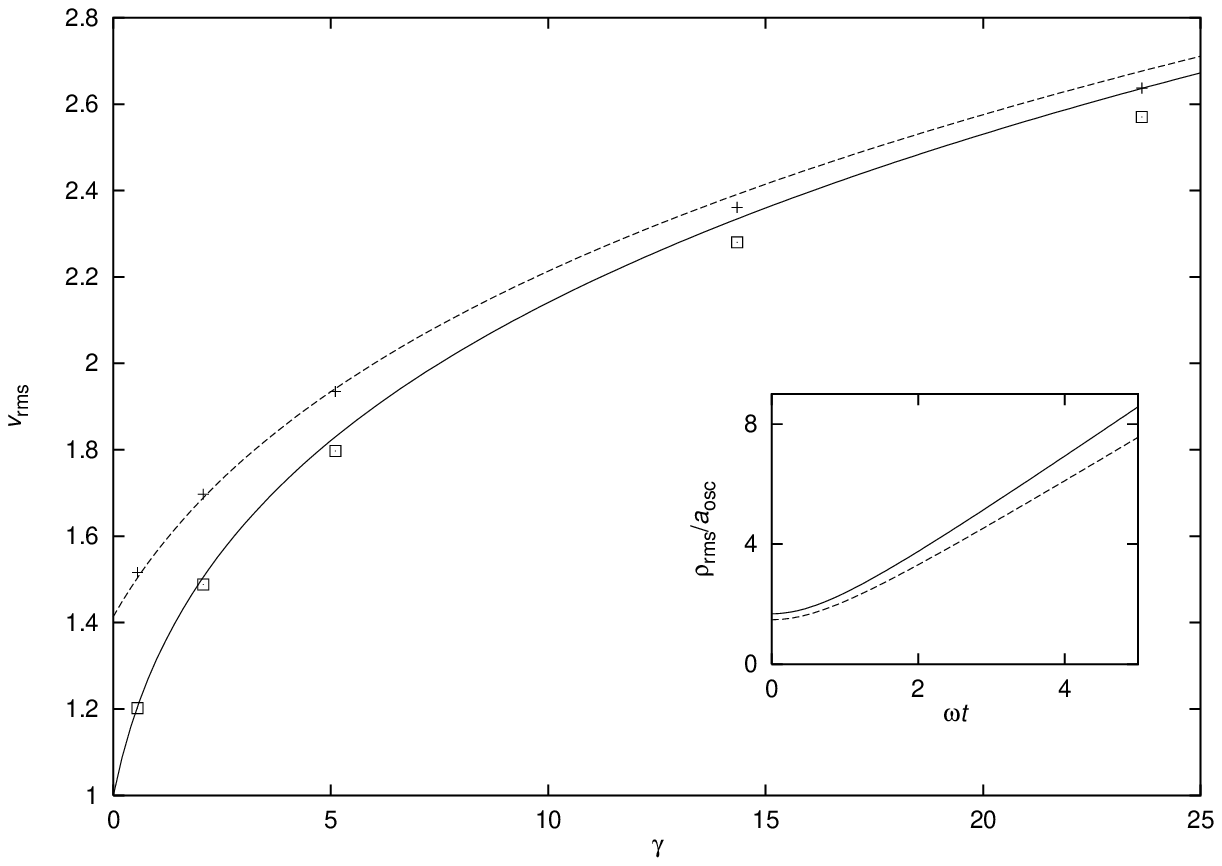,width=11cm,height=8cm,angle=0}
\begin{caption}
{Final root-mean-square velocity for a
two-dimensional cloud in free expansion
as a function of the dimensionless coupling parameter
$\gamma = \nu a$. The full line corresponds to the vortex state in
the variational treatment and the dashed line to the
ground state.
Numerical solutions to the full
time-dependent problem are indicated by crosses (vortex) and squares
(ground state).
The inset shows the time evolution of the rms radius
in units $a_{\mathrm osc}$ for a vortex (full line) and
a ground state  cloud (dashed line), for $\gamma = 2.07$.}
\end{caption}
\end{center}
\label{FIG1}
\end{figure}
As we shall see from the numerical solutions presented below,
the simple result (\ref{fin}) yields an accurate description of
the expanding vortex state.

\subsection{Numerical results}
\label{num2d}
For the numerical study it is convenient to work in terms
of scaled quantities, and we introduce the variables
$f=\psi/ e^{i\phi}(U_0/\mu)^{1/2} $,
$ \rho_1=\rho /a_{\mathrm osc}$
and $\mu_1=\mu/\hbar \omega $. The stationary Gross-Pitaevskii equation
may then be written in the dimensionless form
\begin{equation}
  \label{dsgpe}
  -\frac12 \left(\frac{\partial^2f}{\partial\rho_1^2}
    + \frac1{\rho_1} \frac{\partial f}{\partial \rho_1}
    - \frac{1}{\rho_1^2} f \right)
    + \frac12 \rho_1^2 f + \mu_1 f^3 = \mu_1 f.
\end{equation}
The equation for a non-rotating cloud is similar if
the scaled wave function is defined by
$f=\psi/(U_0/\mu)^{1/2} $, and  the centrifugal
term $f/2\rho_1^2$ (which comes from the phase $e^{i\phi}$
of the  wave function in the case of a vortex state) is omitted.

We integrate this equation using the Runge-Kutta method
to find the wave function $f$ as a function of $\rho_1$ corresponding to a
given dimensionless parameter $\mu_1$. The time-independent wave functions
for $\mu=2.5\hbar \omega$ and $\mu=10 \hbar \omega$ are shown in Fig.\ 2. From
normalization,
we find that these correspond to
$\gamma = 0.55$ and $23.6$, respectively.
\begin{figure}
\begin{center}
\epsfig{file=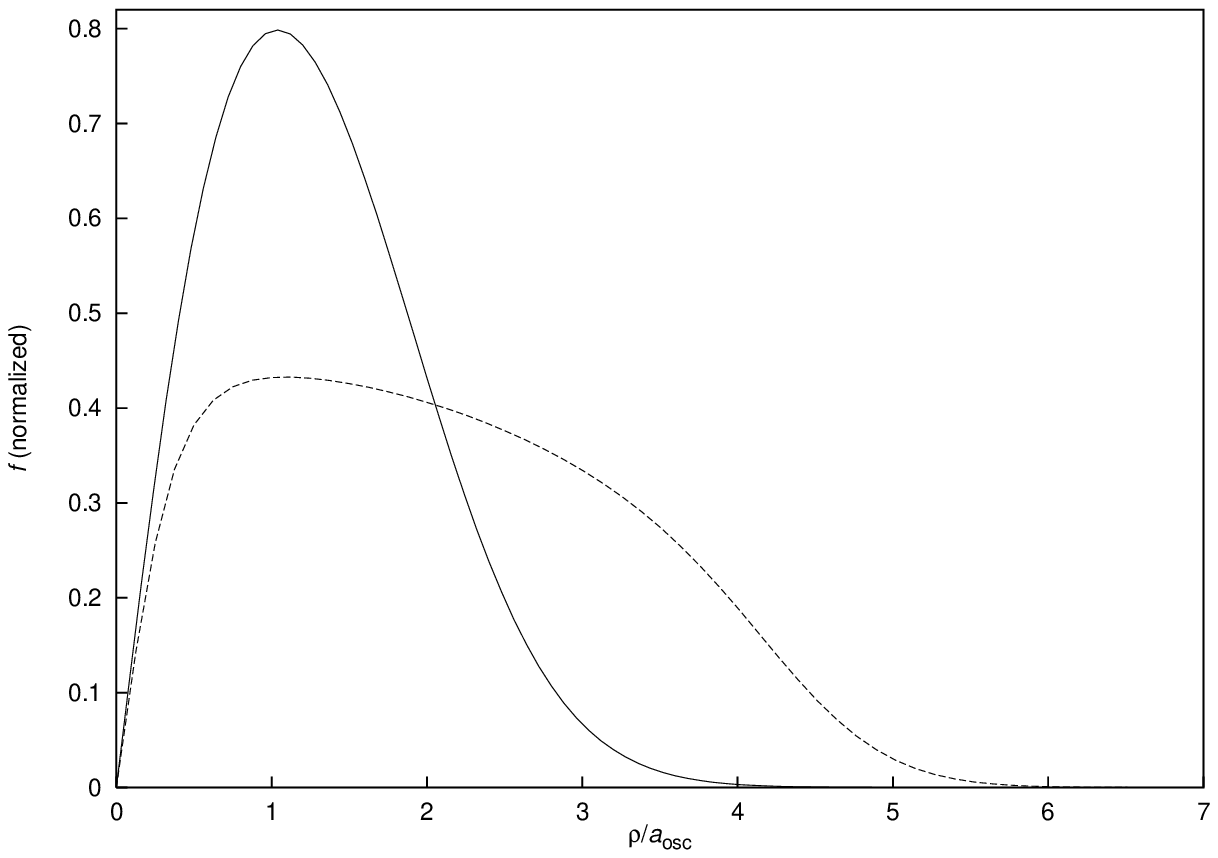,width=11cm,height=8cm,angle=0}
\begin{caption}
{The radial wave function (solution to the
stationary Gross-Pitaevskii equation (\ref{dsgpe}))
for the two-dimensional vortex state. The full line is the
wave function with chemical potential $\mu_1=2.5$, the
dashed line corresponds to $\mu_1=10$. Note that
the units on the $y$-axis
are such that the wave functions are normalized to unity,
i.\ e.\ they are not the same as in Eq.\ (\ref{dsgpe}).}
\end{caption}
\end{center}
\end{figure}
\noindent
To get a feeling for the orders of magnitude, note that for
the realistic case of a system with
an axial dimension of $10a_{\mathrm osc}$
and with $a_{\mathrm osc}/a = 100$, the total number of particles is
$\sim 10^3\gamma$.
In the first experiment performed at JILA \cite{jila1},
$a_{\mathrm osc}/a=700$, $N=2000$ and the  axial dimension
of the cloud was of the
same order of magnitude as $a_{\mathrm osc}$.
In the Ioffe-Pritchard trap at MIT \cite{ioffe},
$a_{\mathrm osc}/a=400$, $N=10^6$ and the length in the $z$-direction is
$\sim 80a_{\mathrm osc}$, giving $\gamma\sim 30$.
The case $\gamma=23.6$ is thus quite close to
typical experimental conditions.

When released from the trap, the wave function evolves according to the
time-dependent Gross-Pitaevskii
equation (\ref{gp}), which in dimensionless form is
\begin{equation}
  i\frac{\partial f}{\partial t_1} =
  -\frac12 \left(\frac{\partial ^2f}{\partial \rho_1^2}
    + \frac1{\rho_1} \frac{\partial f}{\partial \rho_1}
    - \frac{1}{\rho_1^2} f \right)
    + \mu_1 f^3,
  \label{dtdgpe}
\end{equation}
where $t_1 = \omega t$. As before, the non-rotating case is obtained
by dropping the centrifugal term in (\ref{dtdgpe}).
This equation is integrated with respect to time
using the Crank-Nicholson method \cite{crank}, with the initial
value given by the stationary wave functions obtained above.
During the expansion, we compute the rms radius,
$\rho_{\mathrm rms}$,
of the system. We find it to follow Eq.\ (\ref{fin})
to a good degree of accuracy.
At large times, $\rho_{\mathrm rms}$
does indeed grow linearly with time, and the final velocities for
a few different values of $\gamma$ are plotted in Fig.\ 1. The
inset in Fig.\ 1 shows
the evolution of $\rho_{\mathrm rms}$ for $\gamma=2.07$,
corresponding to the chemical potential $\mu=3.54\hbar\omega$
for the vortex state and $\mu=3.07\hbar\omega$ for the ground state.

The Crank-Nicholson method
generates for each time step an explicit table of the real
and imaginary parts of the wave function. On comparing
the wave function at later times with the initial one, we have
been able to find
that the system does indeed undergo a nearly
homologous expansion, i.e. the
wave function does not change its shape, but only flattens
out with time. We shall discuss the physical reasons for such behavior,
in both two and three dimensions, at the end of Section III.

\section{Vortices in three dimensions}

\subsection{Simple variational approach}

The clouds studied in experiments do not possess
the translational invariance in
one direction that was assumed in
our calculations above. It is therefore of interest to
explore the consequences of motion in the third
direction  on the development of a vortex state.
Our discussion for the case of two dimensions may easily
be generalized to three dimensions by employing
the trial function 
\begin{equation}
\psi(\rho,\phi, z, t)= \frac{N^{1/2}}{\pi^{3/4} Z^{1/2}R^{2}}
\rho e^{i\phi} e^{-\rho^2/2R^2}
e^{i\beta\rho^2/2}e^{-z^2/2Z^2}e^{i\gamma z^2/2},
\end{equation}
which describes a cloud undergoing expansion in both
the radial and the $z$-directions.
The result is the two coupled equations
\begin{equation}
\label{eqmotr}
\ddot{R}=\frac{\hbar^2}{m^2R^3}+\frac{NU_0}{4(2\pi)^{3/2}m}\frac{1}{R^3Z},
\end{equation}
and
\begin{equation}
\label{eqmotz}
\ddot{Z}=\frac{\hbar^2}{m^2Z^3}+\frac{NU_0}{2(2\pi)^{3/2}m}\frac{1}{R^2Z^2}.
\end{equation}

The inset in Fig.\ 3
shows the results for the radial and axial expansion
(in units $a_{\mathrm osc}$)
for the case $Na/a_{\mathrm osc} = 20$.
For large times the velocity becomes constant,
just as for the two-dimensional case. In Fig.\ 3, the aspect ratios
of a cloud in an isotropic trap before
and after expansion are shown as a function
of the coupling parameter $Na/a_{\mathrm osc}$. The aspect ratio, $A$,
is defined as the ratio between the root-mean-square
distances, $A=\rho_{\mathrm rms}/(\sqrt{2}z_{\mathrm rms})$,
where the factor of $\sqrt{2}$ takes into account the fact that
there are two coordinates perpendicular to $z$.
With this normalization, the aspect ratio is unity if the rms
values of $x,y$ and $z$ are equal. For the variational wave function
we employ, one finds
$\rho_{\mathrm rms} = \sqrt{2}R$ and
$z_{\mathrm rms} = Z$,
so the aspect ratio is simply $R/Z$.
An aspect ratio differing from
unity in an isotropic trap would be a clear signal for the
presence of a vortex, but from Fig.\ 3 we find that there
are deviations of the aspect ratio from unity
of more than about 5\% only for
a number of particles less than  $100a_{\mathrm osc}/a$.
\begin{figure}
  \begin{center}
    \epsfig{file=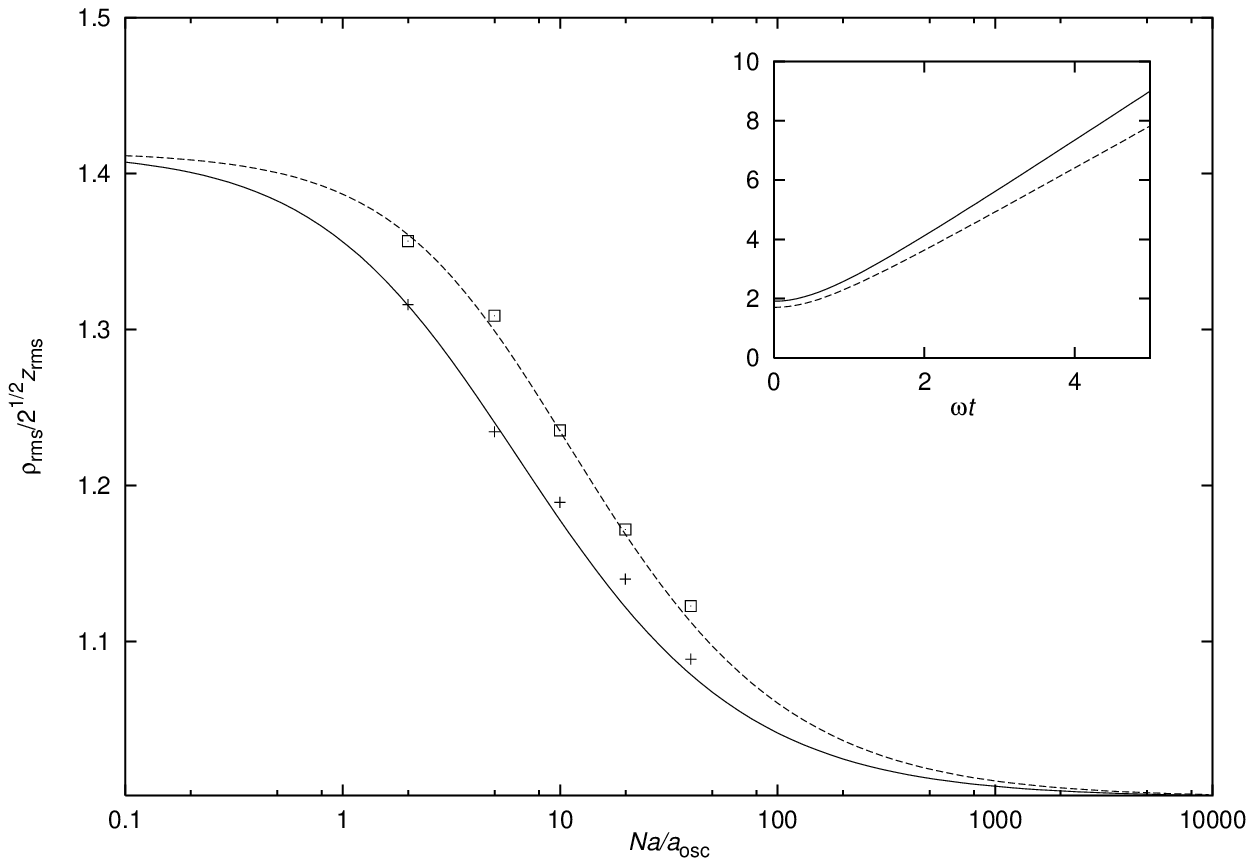,width=11cm,height=8cm,angle=0}
    \begin{caption}
    {Aspect ratio for a vortex system in an isotropic
harmonic trap, before (dashed line) and after (full line)
free expansion. The crosses and boxes are computed using
the semi-numerical scheme described in the text.
The inset shows the expansion for a vortex in the radial (full line)
and axial (dashed) directions, for $Na/a_{\mathrm osc} = 20$.}
\end{caption}
  \end{center}
\end{figure}
\noindent
One may ask whether it is possible to
improve the signal for the presence of a vortex
by changing the external potential. We have therefore computed
the aspect ratio for a system in an anisotropic external potential
$V(\vec{r}\,) = \frac12 m\omega^2(\rho^2 + \lambda^2 z ^2)$ as a function of
the anisotropy parameter $\lambda$. An oblate trap
corresponds to $\lambda >1$, and a prolate one to
$\lambda < 1$. The computation was
carried out for $Na/\bar{a}_{\mathrm osc} = 100$, where
$\bar{a}_{\mathrm osc} = (\hbar/(m\omega\lambda^{1/3}))^{1/2}$
is the geometrical mean of the
oscillator lengths in the spatial directions.
Fig.\ 4 displays the ratio of the aspect ratio
of a vortex, $A_v$, to that of a non-rotating
system, $A_n$, as a function of $\lambda$,
both in the stationary case and after a long expansion
time.
\begin{figure}
\begin{center}
\epsfig{file=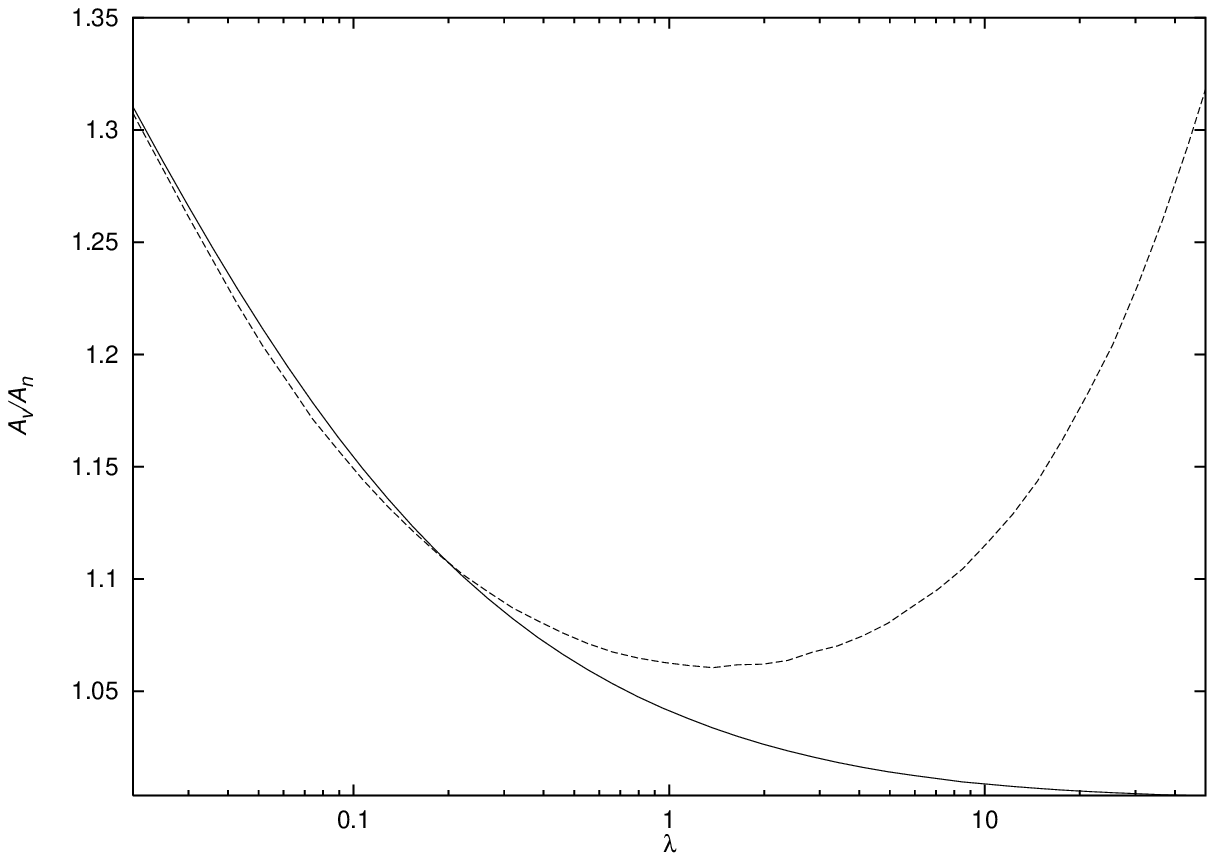,width=11cm,height=8cm,angle=0}
\begin{caption}
{Ratio of the aspect ratio for a cloud containing a vortex, $A_v$, to
that for a nonrotating cloud, $A_n$, in the stationary case (full line)
and after a long time of expansion (dashed line), plotted
as a function of the trap anisotropy
$\lambda$ for $Na/\bar{a}_{\mathrm osc} = 100$.
We see that after expansion the difference in the aspect ratios
is enhanced for a
strong anisotropy, compared to an isotropic trap.}
\end{caption}
\end{center}
\end{figure}
\noindent
We see that for strong oblate anisotropy, the aspect ratio
of a cloud expanding in a vortex state is initially close
to that for the nonrotating state, but following expansion
they are significantly different. For clouds expanding from
a  prolate trap, the ratio of aspect ratios for
the vortex and the nonrotating state  is enhanced both before
and after expansion.

\subsection{Core structure}

In the variational calculations described above, the radial
wave function  is described in terms of a single length scale,
$R$, which determines both the size of the cloud and the
size of the vortex core. This is unrealistic
when particle interactions are important, since the size of
the core is then determined by the
coherence length, $\xi$, which is a function of the density
of particles. To study the dynamics of the core
we employ a more general trial wave function
\begin{equation}
\psi(\rho,z,\phi,t) = \sqrt{N} f(\rho,t) \frac{1}{\pi^{1/4}Z^{1/2}}
        e^{-z^2/2Z^2+i \beta z^2}e^{i \phi}
\label{trial3d}
\end{equation}
with the parameters $\beta$ and $Z$ depending on time. The radial
wave function
$f(\rho,t)$, which is normalized to unity, is to be determined variationally.

Minimizing the action obtained from the Lagrangian (\ref{lagrangian})
with respect to $Z$, $\beta(t)$ and
$f(\rho,t)$ yields the coupled set of
equations
\begin{eqnarray}
  \label{betaeq}
  \beta(t) &=& \frac{m}{2\hbar} \frac{\dot{Z}}{Z},\\
  \label{zeq}
  \ddot{Z} &=& \frac{\hbar^2}{m^2 Z^3} +
        \frac{NU_0}{\sqrt{2\pi}mZ^2}
\left<|f|^4\right>\mbox{, and}\\
  \label{feq}
  i \hbar \frac{\partial f}{\partial t} &=&
    -\frac{\hbar^2}{2m}\left(\frac{\partial^2}{\partial \rho^2} +
    \frac{1}{\rho}\frac{\partial}{\partial \rho} -
    \frac{1}{\rho^2} \right)f
  + \left[ \frac{\hbar^2}{2mZ^2} +
  \frac{NU_0}{4\sqrt{2\pi}Z}\left<|f|^4\right>
  \right]f + \frac{NU_0}{\sqrt{2\pi}Z}\left| f \right|^2 f,
\end{eqnarray}
where $\left<|f|^4\right> \equiv \int d^2\rho |f(\rho)|^4$.
These equations may now be solved numerically, at each
time step simultaneously taking a Crank-Nicholson step for Eq.\
(\ref{feq}) and a Runge-Kutta step for (\ref{zeq}).
To find the initial conditions, we have to
consider the static case with an external potential. We choose to
consider the expansion of a cloud initially confined
in an isotropic potential.
The relevant equations are obtained from Eqs.\ (\ref{betaeq}-\ref{feq})
by neglecting the time derivatives and,
for a potential
$V(\vec{r}\,) = \frac12 m \omega^2(\rho^2+z^2)$,
adding to the right-hand side of (\ref{feq}) the term
$\frac12 m\omega^2\rho^2f$ and to the left-hand side of
Eq.\ (\ref{zeq}) the term $\omega^2 Z$.
Note that when the particle interaction vanishes,
the wave function is that of the lowest state
with unit angular momentum of a particle in the
harmonic oscillator potential.
This corresponds to putting $Z=a_{\mathrm osc}$ and
$f(\rho)=\pi^{-1/2}a_{\mathrm osc}^{-2}\rho e^{-\rho^2/2a^2_{\mathrm osc}}$
in our variational trial function.
Since the Crank-Nicholson method preserves the normalization of the
wave function and $f$ is to be normalized to unity for any value of
the coupling $NU_0$, we may produce any stationary wave function of the system
by starting out with this one-particle wave function and then integrating
the static equations of motion discussed above
with respect to time, while adiabatically
increasing the coupling constant up to the desired value \cite{ruprecht}.
The wave functions we obtain by this method are in good semi-quantitative 
agreement with those obtained by Dalfovo and Stringari 
\cite{dalfovo} from numerical solutions of the three-dimensional  
time-independent Gross-Pitaevskii equation.

Regarding the time development of the  outer radius of the cloud, we find that
the results of the semi-numerical scheme closely follow those found
using the simpler approach, Eqs.\ (\ref{eqmotr}-\ref{eqmotz}),
especially for weak coupling, as the
data points in Fig.\ 3 show.

We have also considered the time dependence of the core size, which
we characterize by the radius, $\rho_i$, at which the particle density
first reaches $e^{-1}$ times
its maximum value. Results for $\rho_i/\rho_{\mathrm rms}$ as a
function of time are shown in Fig.\ 5.
If the radial wave function were of the form of the first excited
oscillator state, $\sim \rho e^{-\alpha \rho^2}$, where $\alpha$ is
a constant, the ratio $\rho_i/\rho_{\mathrm rms}$ would be approximately
0.282.  This is indeed what we find for $Na/a_{\mathrm osc} \lesssim 1$.
For larger values of $Na/a_{\mathrm osc}$ the inner radius is a smaller
fraction of $\rho_{\mathrm rms}$ initially, but the ratio increases with
time. The qualitative behavior of the ratio $\rho_i/\rho_{\mathrm rms}$
with time for $Na/a_{\mathrm osc} \gg 1$ may be understood in terms of
two sorts of processes.
Initially the characteristic time for adjustment of the core size
of the vortex, $\tau_{\mathrm ad} \sim \hbar/nU_0$, is small compared
with the expansion time, $\tau_{\mathrm ex} \sim R/v_s$, where $v_s$ is
the sound velocity of the cloud at $t=0$. Under those conditions
the core of the vortex can adjust essentially instantaneously
to the local density, and thus one expects  $\rho_i$ to scale as the coherence
length, $\rho_i \propto \xi_0 \propto (R^3/Na)^{1/2}$, or
$\xi_0 \sim R(R/Na)^{1/2}$. We take the density $n$  to be $N/R(t)^3$, since
any anisotropy is quite unimportant here as long it is of order unity.
This behavior ceases at a decoupling time, $t_d$, at which
$\tau_{\mathrm ad} = \tau_{\mathrm ex}$. In three dimensions, this means
\begin{displaymath}
R(t) \sim R(0)^{3/4} (Na)^{1/4}
\end{displaymath}
which gives us
\begin{displaymath}
\omega t_d \sim (Na/a_{\mathrm osc})^{1/5}.
\end{displaymath}
Here we have assumed that $R(t) \simeq v_s t$, which holds when
$\omega t \gg 1$. A more careful treatment would result in a somewhat
larger decoupling time.
In two dimensions, $\tau_{\mathrm ad} = \tau_{\mathrm ex}$ implies that
\begin{displaymath}
\frac{R(t)}{R(0)} \sim (\nu a)^{1/2}.
\end{displaymath}
Since $R(t) = R(0)\omega t$ for $\omega t$ large, the decoupling time
is given by
\begin{displaymath}
\omega t_d \sim (\nu a)^{1/2}
\end{displaymath}
in two dimensions.
At larger times the potential energy will play essentially no role,
and the evolution of the cloud will be the same as
for free particles. To the extent that the cloud is expanding homologously
at the decoupling time, we expect that $\rho_i/\rho_{\mathrm rms}$ will
remain constant and equal to its value at the decoupling time. The results
of the numerical calculations exhibited in Fig.\ 5
are consistent with the assumption of homologous expansion at larger times.

We conclude that in two dimensions the expansion is homologous to
a good approximation  during
all of the expansion, while in three dimensions, the relative size
of the core grows until the decoupling time, whereafter it approaches a
constant value.

Our results indicate that by allowing a cloud containing a vortex to
expand freely, the structure will increase in size more rapidly than the
size of the cloud.  In this way one may thus facilitate
optical
detection of the structure associated with the vortex core.

\begin{figure}
\begin{center}
\epsfig{file=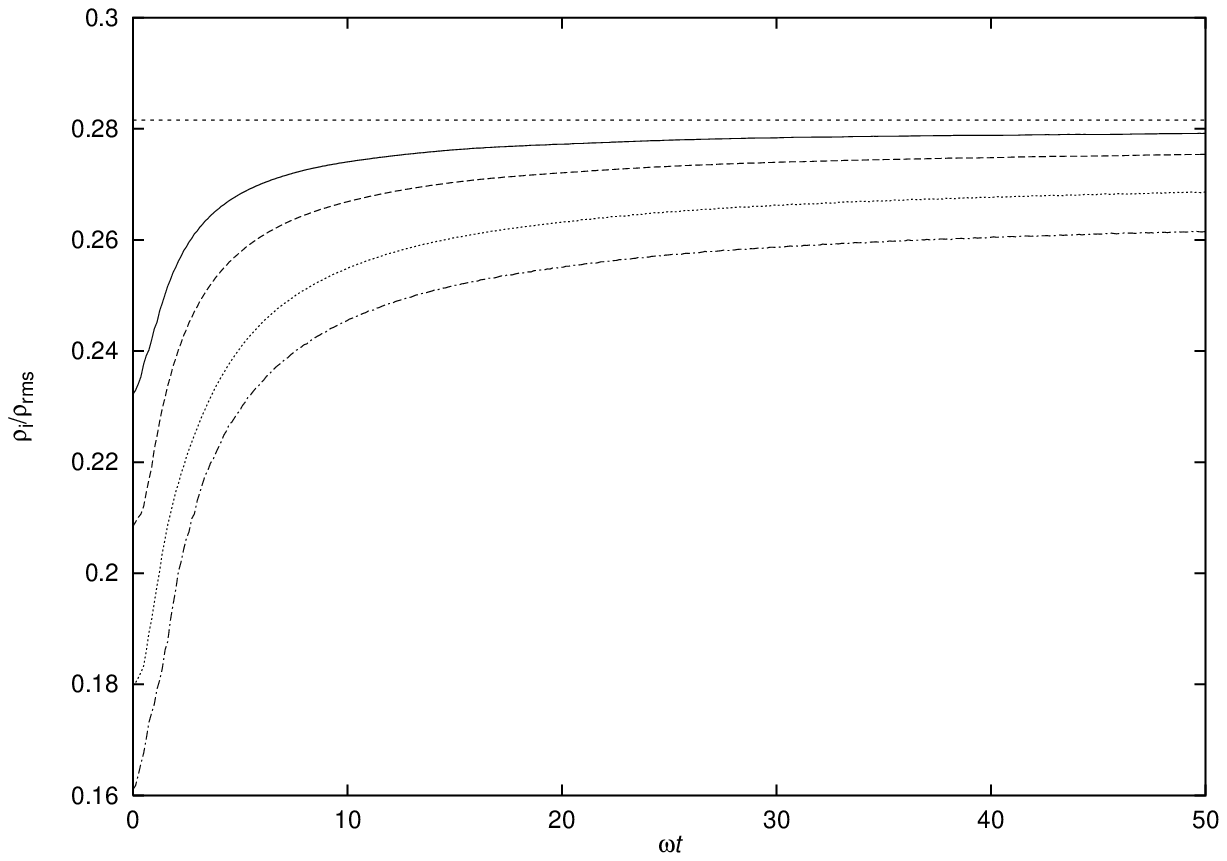,width=11cm,height=8cm,angle=0}
\begin{caption}
{Evolution in time of the ratio of the vortex core radius to
the total system radius for $Na/a_{\mathrm osc}=$5 (full line), 10 (dashed),
20 (dotted) and 30 (dot-dashed), respectively.
The straight line shows the value $\rho_i/\rho_{\mathrm rms}=$0.282,
corresponding to the free-particle limit, $Na/a_{\mathrm osc}=0$.}
\end{caption}
  \end{center}
\end{figure}

\section{conclusions}
In summary, we have described the time evolution of a freely expanding
Bose-Einstein condensed atomic cloud
with a singly quantized vortex in two and three dimensions using both
analytical approximations and direct numerical calculation.
We have found that simple variational estimates describe very well
the time evolution  of the  radius of the cloud.

In the limit of weak coupling, the aspect ratio of the cloud with a vortex
differs from that of the non-rotating ground state both before and after
expansion. This effect may be enhanced by using an anisotropic
trapping potential.

In two dimensions, the cloud is seen to undergo a nearly homologous
expansion, while in three dimensions, the vortex core expands faster than
the size of the cloud during the initial stage of expansion, thus making
it easier to detect experimentally. After a decoupling time $t_d$, estimated
to be $t_d = \omega^{-1} (Na/a_{\mathrm osc})^{1/5}$ for expansion in
three dimensions, the size
of the core will increase linearly with time.

We conclude that a vortex is most easily observed experimentally in the
weak coupling regime, i.\ e.\ for small values of the parameter
$Na/a_{\mathrm osc}$, which is attained either for small
numbers of particles or for shallow traps, since
the relative importance of the vortex to the energy, system size
and velocity of expansion is larger in this case.

\section{Acknowledgement}
One of us (E.\ L.) wishes to thank Nordita for hospitality
during the autumn and winter of 1997-98.

\end{document}